%% file: Main.tex
\documentclass[]{interact}

\usepackage{epstopdf}% To incorporate .eps illustrations using PDFLaTeX, etc.
\usepackage[caption=false]{subfig}% Support for small, `sub' figures and tables

\usepackage[numbers,sort&compress]{natbib}% Citation support using natbib.sty
\bibpunct[, ]{[}{]}{,}{n}{,}{,}% Citation support using natbib.sty
\renewcommand\bibfont{\fontsize{10}{12}\selectfont}% Bibliography support using natbib.sty
\makeatletter% @ becomes a letter
\def\NAT@def@citea{\def\@citea{\NAT@separator}}% Suppress spaces between citations using natbib.sty
\makeatother% @ becomes a symbol again

\theoremstyle{plain}% Theorem-like structures provided by amsthm.sty

\theoremstyle{definition}

\theoremstyle{remark}

\usepackage{longtable}
\usepackage{tabularx}

\begin{document}

\articletype{ARTICLE PUBLISHED at: Journal of Medical Engineering \& Technology (Volume 43, 2019 - Issue 5)}% Specify the article type or omit as appropriate

\title{A Survey of Breast Cancer Screening Techniques: Thermography and Electrical Impedance Tomography}

\author{
\name{J. Zuluaga-Gomez\textsuperscript{$^\ast$, a, b, c}\thanks{$^\ast$ Corresponding author: Juan Pablo Zuluaga, ORCiD: 0000-0002-6947-2706. Prognostics \& Health Management Team, Femto-ST Sciences \& Technologies, Besançon Cedex, 25000. Email: juan.zuluaga@eu4m.eu}
, N. Zerhouni\textsuperscript{a}
, Z. Al Masry\textsuperscript{a}
, C. Devalland\textsuperscript{d}
, C. Varnier\textsuperscript{a}}
\affil{
\textsuperscript{a}FEMTO-ST institute, Univ. Bourgogne Franche-Comté, CNRS, ENSMM, Besançon, France; \textsuperscript{b}Electrical Engineering Department, University of Oviedo, Gijon, Spain; \textsuperscript{c}Universidad Autonoma del Caribe, Barranquilla, Colombia; \textsuperscript{d}Pathology Department, Hospital Nord Franche-Comte, Belfort, France}}

\maketitle

\begin{abstract}
Breast cancer is a disease that threat many women's life, thus, the early and accurate detection play a key role in reducing the mortality rate. Mammography stands as the reference technique for breast cancer screening; nevertheless, many countries still lack access to mammograms due to economic, social and cultural issues. Last advances in computational tools, infrared cameras and devices for bio-impedance quantification allowed the development of parallel techniques like, thermography, infrared imaging and electrical impedance tomography, these being faster, reliable and cheaper. In the last decades, these have been considered as complement procedures for breast cancer diagnosis, where many studies concluded that false positive and false negative rates are greatly reduced. This work aims to review the last breakthroughs about the three above-mentioned techniques describing the benefits of mixing several computational skills to obtain a better global performance. In addition, we provide a comparison between several machine learning technique applied to breast cancer diagnosis going from logistic regression, decision trees and random forest to artificial, deep and convolutional neural networks. Finally, it is mentioned several recommendations for 3D breast simulations, pre-processing techniques, biomedical devices in the research field, prediction of tumor location and size.
\end{abstract}

\begin{keywords}
Breast Cancer; Thermography; Electrical Impedance Tomography; Machine Learning Techniques; Computer Aided diagnosis
\end{keywords}

% Including the required sections for the documet
\input{Chapters/1_introduction}

\input{Chapters/2_thermography}
\input{Chapters/3_electrical_impedance}

\input{Chapters/4_electro_thermal}
\input{Chapters/5_discussion_conclussion}

\section*{Acknowledgement(s)}

This study has been support by the INTERREG institution, section France and Switzerland. Under the framework of the SBRA project.

\section*{Disclosure statement}

The authors have no conflict of interest to declare.

% I did not find this file from the Journal's template, so I commented it. Please uncomment it for obtain the right citation format
\bibliographystyle{tfnlm}

\bibliography{main_bib}

\appendix
\input{Chapters/6_Appendices}

\end{document}

%% file: Chapters/1_introduction.tex
\section{Introduction}

The cancer is a major public health disease that affects many people across the world. The early detection of cancer is mandatory in order to save the patient’s life \cite{1,2,3}. Emerging economies are prone to higher risk of cancer, therefore the socioeconomic factor \cite{4, 5}, aging, unhealthy lifestyle \cite{5,6,7,8} and population growth, may perhaps lead to a higher chance of developing cancer; in addition, the human development index (HDI) is highly correlated with the presence of cancer. In fact, the breast cancer was the first leading cause of cancer-linked death among women in 2018, reaching approximately 15\% of the total number of registered cancer deaths \cite{9}. Mammography, ultrasound and magnetic resonance imaging, stand as the main techniques for breast cancer screening, however, limitations like x-rays, expensiveness and misinterpretations have let to grow in popularity alternative techniques such as thermography and electrical impedance tomography (EIT).

The Globocan’s 2018 fact sheet from the International Agency for Research on Cancer - World Health Organization (WHO) shows the number of new cases and deaths in 2018 from cancer. During 2018 the male number of cancer's new cases reach more than nine million, and more than eight million of new females where registered as well. Globocan and other authors \cite{4, 9} have predicted a 50\% rise in the number of prevalence in breast cancer by 2040. Indeed, the proportion of breast cancer deceases may vary depending in the world’s region. Specifically, studies have uncovered that the breast cancer mortality-to incidence ratio in developed countries is 0.20, where in less developed countries is 0.37, almost twice \cite{4, 7}.

Early detection of breast cancer could increase the survivability rate up to 90\% of all cases within a five-year window, thus, the needed of an easy-access, cheap and trustable breast cancer screening method still latent in underdeveloped countries \cite{4, 7}. In the other hand, some countries keep multiple barriers for establish an effective breast cancer screening system, e.g., organizational, psychological, structural, sociocultural and religious \cite{10}. To exemplify, the Kaiser Family Foundation in late 2018 have reported that 11\% from the total amount of women in USA have not any kind of social insurance, which represents more than 10 million women \cite{11}. Differently, a few countries have religious rules where the woman cannot expose the breast; therefore, the reference methods are non-viable for an accurate and prior detection of breast cancer. In contrast, devices and techniques that would not need physicians’ direct contact like thermograms or bio-impedance images will make a considerable impact. Presently, several techniques are available in the medical field for breast cancer screening, despite the variety, the main differences lie on cost, method, specificity, sensitivity and patient’s discomfort during test. The table \ref{x_comparison_between_technique} show a comparison of the main techniques for breast cancer diagnosis and screening described by Kandlikar et al. \cite{12}. 

Mandelblart et al. \cite{13} have mentioned that the early spotting of an abnormality could reduce between 15 to 25\% the mortality index rate, \cite{14} also obtained similar results. The Mammography is the main techniques for breast cancer screening and it is based on x-rays. In spite of the mammograms' benefits, the over-diagnosis (false positives), painful procedure, high number of false negatives (usually when the person who evaluate the results, make erroneous assumptions, or in dense breast) and use of x-rays have been making it a method which need to be renovated \cite{15} or even replaced by state-of-the-art techniques like thermography and EIT. Under those circumstances, no matter the individual risk of breast cancer, either, genetically (family) or unhealthy lifestyle, the current guidelines suggest breast checks every 1 or 2 years starting at age of 40 or 50 years \cite{13}. 

In general, more information about, guidelines, health benefits, recommended gap time between tests and type of breast cancer, are in \cite{13, 16, 17}. Truthfully, the European Commission has published a document regarding the breast cancer screening and diagnosis guidelines, summarizing that an accurate system is made of different steps as follows: (i) screening, (ii) diagnosis,  (iii) communication to the patient, (iv) interventions to reduce inequalities, (v) monitoring and (vi) evaluation of the tests. \\

"Insert table \ref{x_comparison_between_technique} here (the table must be in landscape mode, therefore, it should be let on appendices)" \\

% setting the counter for the new tables 
\setcounter{table}{1}

A wider explanation in breast cancer techniques for diagnosis is in Warner's report \cite{18}. The need for cheap, effective and free side effects breast cancer screening techniques have led the development of several new techniques like thermography and EIT, henceforth, this article will review their last breakthroughs.  

\textbf{Given these considerations, thermography, infrared imaging and electrical impedance tomography have emerged as cheap and reliable solutions for breast cancer diagnosing.} Firstly, the thermography is the measurement of the skin’s temperature, depending on the methodology, the results could vary. For example, if an infrared camera (IR) is used, the category is infrared imaging, on the other side, if the method employs either, sensors attached to the region of interest (ROI) or liquid crystal thermometer the method is simply thermography, producing a temperature matrix. Consequently, many researchers have found a huge correlation between the increase of heat and blood perfusion rates in tissue surrounded by a tumor, indeed, higher than normal tissues. Secondly, EIT is a technique which evaluate the inner electrical conductivity or impedance (resistance) distribution of a given body; the signals are collect with electrodes in contact with the skin’s ROI \cite{19, 20}. To clarify, similar to the increase in the temperature of cancerous tissue’s surrounds, the malignant tissue has more than twice times higher impedance than the normal one. The majority of physical, electrical and thermal characteristics possess differences \cite{21, 22}. In addition, many authors have presented several EIT systems for breast cancer diagnosis \cite{12, 19, 20, 21, 22, 23}. 

The so called Computer Aided Diagnosis (CAD) system, are computational algorithms capable of identify patterns in almost whichever type of data; now, several research teams are struggling to add the CAD systems in the diagnosis phase in order to decrease the false positive (FP) and false negative (FN) rates. Patricio M., et al give insights in this type of systems, where a human expert and a CAD system work together in order to give better results than standalone systems \cite{24}. Normally, a CAD environment is made of a five-step pipeline, including identification, data preprocessing, feature extraction, prediction or classification, and post-processing.

The paper is organized as follows. Section 2 conveys the last breakthroughs regarding thermography such as main protocols, 3D breast simulations and machine learning approaches for thermal databases. Similarly, Section 3 is focused in electrical impedance tomography. Section 4 covers the promising progress in two-steps systems, mixing electrical impedance tomography and thermography for performance improvement. Finally, Section 5 and 6 formulate the discussion, conclusion and comments about future works. 

\nopagebreak

%% file: Chapters/2_thermography.tex
% Second chapter
\nopagebreak
\section{Thermography}

Thermography is the measurement of the temperature based on infrared radiation. In contrast to other modalities, it is a non-invasive, non-intrusive, passive and radiation-free technique. In medicine, the skin's surface temperature exposes many features because, human skin's radiance is an exponential function of the surface temperature, in other words, is influenced by the level of blood perfusion in the skin \cite{25}. In fact, Krawczykm et al. summarize "Thermal imaging is hence well suited to pick up changes in blood perfusion which might occur due to inflammation, angiogenesis or other causes" \cite{26}. As mentioned before, the early detection of breast cancer provides significantly higher chances of survival \cite{3, 27}. Thermography has advantages over other techniques, in particular when the tumor is in an early-stage or in dense tissue \footnote{Dense tissue: high index of fibrous or glandular tissue and low of fat} \cite{28}. Many authors\footnote{AACR, American Association for Cancer Research} had explained before the high risk of developing breast cancer where is high mammographic density \cite{29}, also in \cite{30}  demonstrated the correlation between body weight, parity, number of births and menopausal status, regarding to breast cancer. The above authors have point out the highly rate of mammograms' false positive cases and the fact that mammography can detect tumors only once they exceed certain size. Nonetheless, thermography could be a solution to these problems. In the medical field, diagnostic of breast cancer using thermography keeps having two different points of view, one side explain that thermography images produce a high number of false positives, e.g. the thermal images were not enough for the initial evaluation of symptomatic patients in Kontos research \cite{31}. Similarly, some authors mention low precision and recall \cite{32, 33} after the initial evaluation. On the other hand, other authors explain that the progress in computational tools, machine learning and infrared cameras, could situate the thermography as a technique capable of overcoming the limitations of mammography. 

\subsection{Initial years of thermography}

The first time ever that was used a thermal/infrared imaging to breast cancer diagnosis aid was in Montreal in 1956 when Lawson \cite{34} recorded the skin's heat energy using a "thermocouple". The thermocouple is a device made of two dissimilar metals that allows calculating the electromotive force created by the juncture of two metals \cite{34}. He mentions that Massopoust and Gardner had used a system called "Infrared phlebogram \footnote{(1) A graph indicating the pulsing of the blood within the vein. (2) An X-ray image of a vein that has been injected with a dye that is visible on the image taken, Collins Dictionary}" to aid the diagnosis of breast complaints \cite{35} in 1200 cases. Nevertheless, not was before 1958 when Lawson presented one of the first devices capable of create an infrared imaging. He described the process as follows; "At any instant during the scan, the infrared energy radiated from the point on the body at which the scanning mirrors are "looking", is reflected on to a parabolic mirror, thereby focusing the energy from a point on the object on the infrared detecting cell" \cite{36}. The infrared imaging device was called "Thermoscan", but not was before 1965, when Lawson's team obtained the first patent regarding a thermal device as a diagnostics tool \cite{37}. 

Afterwards, a team from Texas used a device called Pyroscan for measure the skin temperature. They considered that the equipment was expensive but technically simple, however, the false positives rates were similar to mammography tests \cite{38}. Williams et al. likewise present studies with many common features. In 1960 \cite{39} and in 1964 they were granted a patent \cite{40} of a device for measure the skin temperature. On the other hand, Mansfield et al. carry a research, testing different heat-sensing devices in cancer therapy to contrast methodologies \cite{40_1}. Swearingen in 1965 concluded two main things, first, the true positives rates was greatly increased when mammography and thermography were applied together, second, the thermography was seen as a new technique for diagnostic procedure in mass screening of the breast \cite{41}. During the 20th century, was conceived a patent \cite{42} using an infrared radiometer mounted on a carriage guided path; \cite{43} patented the process of diagnosis a disease through thermography. In 1971 Isard et al. cooperate in a ten-thousand-cases study. During the four-year research they determined that 61\% of cases were correctly diagnosed with thermography, 83\% with mammography and 89\% applying simultaneously both techniques \cite{44}. Isard et al. \cite{45} studied the pathological changes in spatial distribution of temperature over the skin surface.

\subsection{Protocols for thermography}

The thermography test could be considerably affected when guidelines are not followed. Throughout the last years, many studies lacked standards and protocols when record thermograms; those could be one of the primary reasons for the poor results. Ng \cite{46} and Satish \cite{12} mention several standards to follow, in order to obtain high quality and unbiased results. Firstly, it is recommend that patients should avoid tea or coffee before the test, large meals; alcohol and smoking may affect the physicist’s or CAD’s judgment. Secondly, the camera needs to run at least 15 min prior the evaluation, keeping a resolution of 100mK at 30 $^{\circ}$C at the same time the camera should have at least a 120x120 points temperature matrix. Third, is recommend a room’s temperature between 18 and 25 $^{\circ}$C, humidity between 40\% and 75\%, carpeted floor and avoid any source of heat. The post-processing phase should be able of identify the type of breast cancer, either, made by a physicians or CAD system. Similarly, Ng et al. in a ninety patients study propose a temperature-controlled room between 20$^{\circ}$C and 22$^{\circ}$C with and humidity of 60\% $\pm$5\%, the patient rested for 15 minutes\cite{tem_new_1}. On the other hand, in order to ensure that patients are within the recommended period, they needed to be in the 5th to 12th and 21st day after the onset of menstrual cycle, since at this time the vascularization is at basal level, with least engorgement of blood vessels \cite{tem_new_2}.

\subsection{Temperature-based technologies for breast cancer diagnosis}

The term “thermography” is not limited to measure the skin’s temperature, but also rearrange these values in one “image”, like an illustration, creating a heat map of the breast’s ROI, where each “pixel” express an equivalent temperature value. Ng et al. mention that the presence of localized or focal areas of approximately 1.0$^{\circ}$C or more (including the areola region) and significant vascular asymmetry forming "clusters" are features that need to be considered as abnormal \cite{tem_new_1}. They obtained an global accuracy of 59\%, and true positive accuracy of 74\% using Bayes Net. Arena et al. \cite{47} in 2003 mention the benefits of the digital infrared imaging also called "DII". They tested a weighted algorithm in 109 tissue proven cases of breast cancer, generating positive or negative evaluation result based on six features (threshold, nipple, areola, global, asymmetry and hot spot), they employed an infrared camera with a 320x240 pixels (temperature points), and sensitivity of 0.05 degrees. Contrary, some authors mention that the static temperature measurement could carry out mislabeling during the evaluation process, for that reason they adviced to use a “dynamical” approach instead. Comparatively, some researchers are not only focused on breast cancer labeling, but rather in the localization and size itself of the tumors. Partridge and Wrobel modeled in 2007 a method using dual reciprocity coupled with genetic algorithms to localize and size breast tumors; nonetheless, they concluded that smaller and deeply located tumors, produce only a limited perturbation making then impossible to be detected \cite{48}; The estimation of tumor characteristics can be found in \cite{49}. Kennedy et al. compared the advantages and disadvantages of thermography, mammograms and ultrasound. Arena et al. \cite{50} specified that thermograms are early indicators of functional abnormalities that could lead to breast cancer.

The infrared cameras used for thermography provide the result in both, a temperature matrix or a heat map image. Rajendra et al. \cite{51} built an algorithm using support vector machines classifier for automatic classification of normal and malignant breast cancer instances, they used the database made by \cite{52, 53}. Later, in 2009 Schaefer et al. performed a fuzzy logic classification algorithm having an accuracy of nearly 80\%, with a population of 150 cases; they explain that statistical feature analysis is a key source of information in order to achieve a high accuracy, i.e., symmetry (mean) between left and right breast and standard temperature deviation \cite{54}. Araujo presented a symbolic data analysis on 50 thermograms (data type: temperature matrices), obtaining four variables, minimum and maximum temperature values from the morphological and thermal matrices; during the training process was implemented leave one out cross validation framework for reduce the over-fitting \cite{55}. Marques and Da Silva \cite{52, 53} provide the only public database for thermography. Nevertheless, it was not after 2015 when \cite{53} propose a system to diagnostic breast cancer among a population of more than 50 patients. A recent study from Da Silva et al. present a renovated database, with a web-page encompassing all the cases with the corresponding validation technique (mammography or ultrasound) \cite{53_1}.   

\subsection{Computer aided techniques in thermography}
\label{ch:CAD_techniques}
Computer-aided diagnosis (CAD) has become one of the major research topics regarding medical imaging, these systems assist physicians in the interpretation of medical images such as, x-rays, MRI, infrared imaging and ultrasound. A big part of the studies related with CAD systems and infrared imaging techniques for breast cancer diagnosis employ the public database from \cite{52, 53, 53_1}. Nevertheless, some authors have created non-public databases that are used for private purposes only. Ng et al. \cite{tem_new_1} presented a computerized detection system with bayes net rules on a ninety patients group, the algorithm yield a 59\% accuracy, then in 2002 they proposed a new system using artificial intelligence for detect breast cancer. Similarly, Ng's team \cite{tem_new_3, tem_new_4} employs an artificial neural network (ANN) coupled with a bayes net ruler obtaining an accuracy of 61.54\%, but not was before 2008 when his team create a two-steps algorithm, where a linear regression decided whether to choose a ANN with radial basis function or a back-propagated ANN achieving a 81\% accuracy. Several authors have been using ensemble machine learning (ML) systems to yield a better performance, e.g. Mambou et al. \cite{2} developed a Deep Neural Networks (DNN) and support vector machines algorithm using the mentioned before database. Initially, they pre-processed each thermal image for fitting them in a DNN, then, they extract and normalize the features for feeding them into a ML algorithm. The database is composed of 56 Brazilian patients where 37 had anomalies and 19 were healthy women. The last decade's improvement in microcontrollers and personal computers allowed the propagation of software in Machine Learning Techniques (MLT), such as Python, Matlab, Orange3 (based on Python) and WEKA. Therefore, day-to-day there is more researchers working on MLT and medical imaging.

Nowadays there is four groups of MLT distributed as follows: (i) supervised, (ii) unsupervised, (iii) semi-supervised, and (iv) reinforcement algorithms. The current breast cancer detection systems are grouped in supervised and unsupervised. Firstly, we recall the supervised algorithms like linear and logistic regression, linear discriminant analysis (LDA), Gradient Boosted Trees as AdaBoost (AB), Support Vector Machines (SVM) with kernels (like, Radial Base Function - RBF, or Gaussian), Naive Bayesian Networks (NBN), Decision Trees (DT), Random Forest (RF), Artificial Neuronal Networks (ANN) and Deep Neuronal Networks (DNN). Secondly, K-nearest neighborhood (KNN), Principal Component Analysis (PCA), locally-linear Embedding and linear discriminant analysis stand as unsupervised algorithms. Those techniques usually make part in big artificial intelligent environments, which have several types of biomedical databases (e.g. thermal images). Even though the number of instances or population size is a key factor to develop a robust MLT, in some cases the volume of the database is not a drawback, rather the dataset’s balance. As a result, Krawczyk et al. in 2013 proposed an ensemble algorithm \footnote{Meta-algorithms that combines several machine learning techniques into just one predictive model decreasing variance, bias and accuracy, the resulted model is better than the other ones separately.} for clustering and classification in a small and unbalanced database of breast thermal images. They implement a K-fold cross-validation to reduce the bias and over-fitting of the model. It works as follows, first the database is divided in training and testing set; the training set is split in "K" number of folds (i.e. five folds). Finally, the model is trained five times, where one fold acts as evaluation and the other four ones act as training; the process is rotated over all folds \cite{25} and the results is averaged. 

\begin{figure}[htb]
\centering
\includegraphics[width=5.5in]{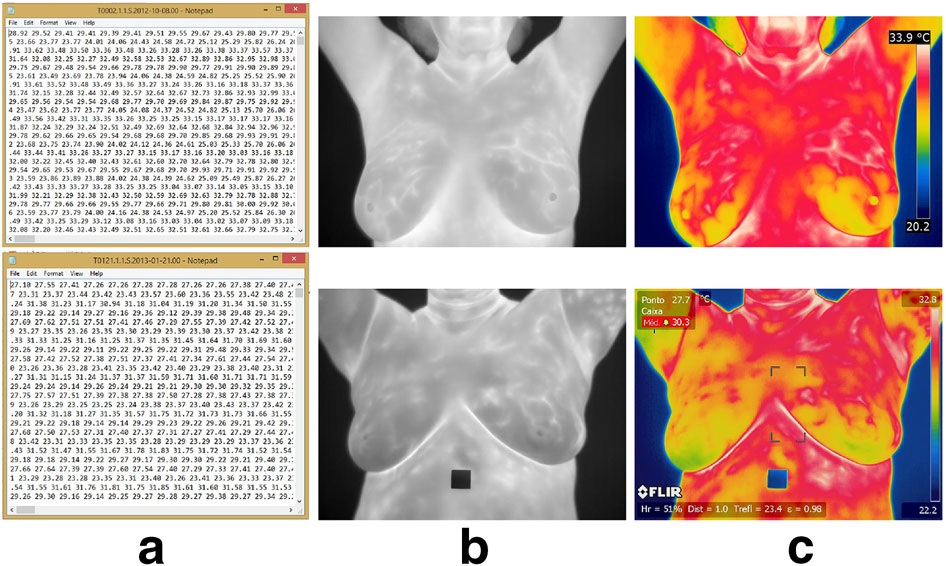}
\caption{Representation of breast thermograms (a) Temperature matrix (b) Grayscale image (c) Pseudo-color image from \cite{59}.}
\label{thermal_representation}
\end{figure}

During, the last six years, several reviews \footnote{In \cite{56} is mention that 5\% of the all articles from January 2012 to January 2017 regarding CAD techniques for diagnosis of breast cancer mention thermography as main method.} regarding infrared technologies have emerged, establishing with it a well-delimited guidelines of the main protocols and mew directions of breast cancer diagnosis \cite{12, 56, 57}. The region of interest pre-processing of thermal images is another issue to manage in order to improve the global performance of the ML algorithms. \cite{58} mentioned an optimized method breast thermal images breast segmentation using extended hidden Markov models (EHMM), bayes net and random forest in a 140-indivuals database from the IUT OPTIC (non-public database) from Iran. Furthermore, Sathish et al. explained that the thermal camera's information can be interpreted in three ways. Temperature matrix, gray scale image and pseudo-color image (or heat map), being the last, the one which possesses more information than the first two. They concluded that the normalization (either, max-min or mean-std normalization) of thermal images could improve the algorithm's performance \cite{59}. Furthermore, the Figure \ref{thermal_representation} show the three possible variations of thermal images representation. Surely, the constant (i) increasing in computing performance, (ii) the price reduction of microcontrollers and (iii) the increase of breast cancer among women, have brought more and more researchers interested in non-conventional techniques for an accurate diagnosis. 

\textbf{Over the last few years some researchers have focused more in dynamic processes rather than static ones. For example, Silva et al. \cite{60} propose 39 supervised and unsupervised MLTs using dynamic infrared thermography rather than the common "static" approach. This methodology quantitatively estimate the temperature changes on a given surface, after a thermal stress. Therefore, their classification algorithms yielded an average accuracy of 95.38\%. Further information about the protocol could be found in \cite{52, 53, 53_1}. Gonzalez-Hernandez et al. \cite{61} reviews the potential of dynamic thermography and the possible future directions for breast cancer diagnosis. On the other hand, Joro et al. \cite{tem_new_5} concluded that image processing before frequency analysis affect the outcome of dynamic infrared imaging in breast cancer, but the mammographic location of the cancer. Finally, Mambou et al. \cite{2} proposed an ensemble methodology using DNN and SVM. \cite{62} developed a intelligent textile to measure skin temperature that could be useful for further dynamic thermography studies rather than thermal cameras.}

Table \ref{summarize_table} summarizes the main references regarding thermography and CAD systems for breast cancer diagnosis. The first column comment the scope of the project and the main methodology implemented. The second one indicates which machine learning technique is used in order to predict the breasts status. The last column exhibits the main achieve results.

\begin{longtable}{p{7cm}p{3.2cm}p{2.3cm}c}
\caption{Summarized thermography methods. The main parameters of evaluation are: Accuracy (Acc), sensitivity (Sen), specificity (Sp), AUC (area under the curve), ROC (receiver operating characteristic curve) and PPV (positive predictive value).}
\label{summarize_table}
\toprule
    Scope of the project &  Machine Learning \newline Technique (MLT) & Evaluation \newline Result &  Ref. \\
\midrule
	
	Clustering and selection using several MLT in one ensemble unit, 5 folds cross-validation & SVM-RBF \newline DT \newline RF & Acc: 90\% \newline Sen: 82.6\% \newline Sp: 91.9\% &\cite{25}\\ 
	
	First statistical approach for breast thermograms in a ninety patients study, using a 256x200 pixels IR camera & Bayes Rules & Acc: 59\% \newline Sen: 54\% \newline Sp: 67\% \newline PPV: 74\% &\cite{tem_new_1}\\

     Classifying normal or at risk in a previously confirmed database which cancer, normal and post-operation patients& No MLT  \newline Weighted Algorithm & Acc: 99\% \newline Sen: 99\% & \cite{47}\\
     
     Localization of skin tumors, based on temperature & Genetic algorithm & Acc: 100\%\footnotemark &\cite{48} \\
     
     Presence, location, size and properties of the tumor & Genetic algorithm & Max Error \newline E:2.6\% &\cite{49} \\
          
     Analysis and comparison of thermography vs breast cancer screening techniques & No MLT & Acc: 83\% \newline Sen: 83\% & \cite{50}\\
     
     Thermography classifier of breast cancer. They provide a user-friendly graphical user Interface & SVM & Acc: 88.1\% \newline Sen: 85.7\% \newline Sp: 90.48\% &\cite{51}\\   
     
     Statistical features from infrared signals and asymmetry from both breasts & Fuzzy Logic classifier & Acc: 79.5\% \newline Sen: 79.9\% \newline Sp: 79.5\% &\cite{54}\\

	Symbolic data analysis for classification as malignant, benign and cyst breasts thermograms from \cite{53, 53_1} database & Linear Discriminant \newline Parzen-window & Acc: 84\% \newline Sen: 85.7\% \newline Sp: 86.5\% &\cite{55}\\
		 
	AI approach for breast cancer diagnosis with thermograms. They propose an optimal time gap during the menstrual cycle to perform the test & ANN, Bayes Rules & Acc: 61.5\% \newline Sen: 69\% \newline Sp: 40\% \newline PPV: 90.91\% &\cite{tem_new_3}\\
	
		Integrated technique for breast cancer diagnosis using a bio-statistical method as pre-processing technique & ANN, linear regression (RBFN) & Acc: 80.9\% \newline Sen: 100\% \newline Sp: 71\% &\cite{tem_new_4}\\

	Current status of breast cancer diagnosis with thermography \cite{2, 57, 12} & Many MLT & -- & \cite{56}\\
	
	Breast thermal images segmentation as a pre-processing method & EHMM & Execution time reduced &\cite{58}\\ 		

	Normalization of breast cancer infrared images improve the global accuracy (min-max approach)& SVM \newline Kernel: Gaussian & Acc: 91\% \newline Sen: 87.23\% \newline Sp: 94.34\% &\cite{59}\\ 		

	Analysis of thermal breast images as time series. First, region of interest (ROI) is segmented and then \textit{k-means} is implemented & Bayes Net \newline Decision Table \newline RF & Acc: 95.4\% \newline Sen: 95.37\% \newline Sp: 95.4\% \newline ROC: 0.97 &\cite{60}\\ 
	
	Extensive review in last advances regarding dynamic breast thermography's & Many MLT&  --  &\cite{61}\\
		
	Analysis of thermal patches as key features for determine the presence of cancerous tissue in the breast & SVM, DT, AB, RF, KNN, ANN, LDA& Acc: 98\% \newline Sen: 98\% \newline Sp: 98\% &\cite{66}\\

	Thermal breast model in COMSOL\textregistered , with tissue properties per layers & -- & -- &\cite{64}\\ 
			
	Tumor localization from skin temperatures in COMSOL\textregistered  & -- & -- &\cite{65}\\ 
			
	Comparison between Temperature based analysis, intensity based analysis and tumor location matching & SVM & Acc: 83.2\% \newline Sen: 85.6\% \newline Sp: 73.2\% &\cite{66_1}\\		

	Analysis of thermography (DITI) in the diagnosis of breast mass, side diagnostic with Ultrasound and/or MRI as validation techniques& No NLT & Acc: 79.6\% \newline Sen: 95.2\% \newline Sp: 72.8\% &\cite{63}\\	
\bottomrule
\end{longtable}

\footnotetext{Tiny tumors cannot be detected with this method \cite{48}}

\subsection{Breasts model based on 3D simulation and thermal properties}

\begin{figure}[htb]
\centering
\includegraphics[width=3in]{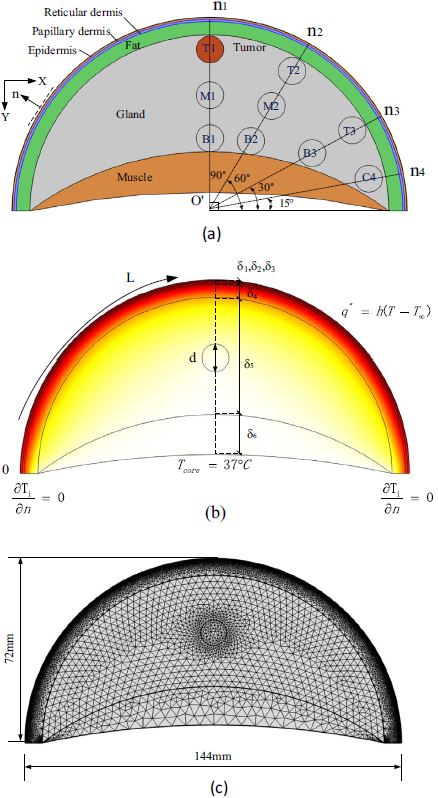}
\caption{(a) Schematic of the breast tissue layers and the tumor locations in a computational domain; (b) schematic of the breast tissue layers’ dimension with boundary conditions for steady state; (c) the computational mesh and breast tissue dimensions, from \cite{64}.}
\label{thermal_3D_model}
\end{figure}

The temperature emanated from a human breast may vary depending on a range of features, both, static and dynamical. The first are tumor size, depth and location, volume of the breast and quadrant of the suspected tumor. On the other hand, the pathophysiological characteristics are different from patient to patient, therefore, some authors have implemented DITI, where the breast undergo a thermostimulation reducing the surface temperature, then it is let to reach a steady state temperature (the temperuture is measured during the test). The review from Zhou and Herman \cite{64} present 3D models of the heat distribution in healthy and non-healthy breasts. The Figure \ref{thermal_3D_model} depicts a breast 3D model in COMSOL\textregistered ~for computing the heat distribution when a tumor is present; \cite{65} present similar results. An analysis of thermal patches in the breast could improve many algorithms accuracy \cite{66}, also Gogoi et al. \cite{66_1} propose a method to locate suspicious regions in thermograms matching them with tumor locations in mammograms, thus, knowing the ground true, they were able to evaluate the efficiency in 3D models and real thermal images.   

\begin{table}[htb]
\centering
\caption{Thermal properties of breast tissue layers (from \cite{64})}
\label{temp_properties}
\begin{tabular}{l p{1.3cm}p{1.5cm}p{2cm}p{1.3cm}p{1.5cm}p{1.5cm}}
\toprule
& \multicolumn{2}{l}{Properties} \\ \cmidrule{2-7}
Tissue layer & \small Thickness $\delta ~(mm)$ & \small \raggedright Specific heat C $(J/Kg~K)$ & \small \raggedright Thermal conductivity k(W/m K) & \small  Density $\rho$(kg/$m^3$) & \small Perfusion rate $w_b~(1/s)$ & \small Metabolic HG Q(W/$m^3$) \\
\midrule
Epidermis & 0.1 & 3589 & 0.235 & 1200 & 0 & 0 \\
\raggedright Papillary dermis & 0.7 & 3300 & 0.445 & 1200 & 0.00018 & 368.1 \\
\raggedright Reticular dermis & 0.8 & 3300 & 0.445 & 1200 & 0.00126 & 368.1 \\
Fat & 5 & 2674 & 0.21 & 930 & 0.00008 & 400 \\
Gland & 43.4 & 3770 & 0.48 & 1050 & 0.00054 & 700 \\
Muscle & 15 & 3800 & 0.48 & 1100 & 0.0027 & 700 \\
Tumor & d=10 & 3852 & 0.48 & 1050 & 0.0063 & 5000 \\
\bottomrule
\end{tabular}
\end{table}

Pennes in 1948 \cite{67} proposed the Equation \ref{Pennes_equ} as a heat transfer model of the human tissue:

\begin{equation}
\label{Pennes_equ}
\rho _i c_i \dfrac{\partial T_i}{\partial t} = k_i \nabla ^2 T_i + \rho _b c_b w_{b,i} (T_b - T_i) + Q_i
\end{equation}
 where $i$ represents the breast tissue layers of epidermis, papillary dermis, reticular dermis, fat, gland and muscle respectively. $\rho _i$, $c_i$, $k_i$, $T_i$, $Q_i$ and $w_{b,i}$; correspond to tissue layer density, specific heat, thermal conductivity, temperature, metabolic heat generation (HG) rate and blood perfusion rate, respectively. Then, $\rho _b$, $c_b$ and $T_b$; stand for blood density, blood specific heat and arterial blood temperature, respectively. also called, a transient heat conduction Bioequation (\ref{Pennes_equ}), helped \cite{64} research to develop 3D models with the properties of the Table \ref{temp_properties}. 

A last key point to realize is the comparison between steady state and dynamical thermography. As mentioned before in section \ref{ch:CAD_techniques}, while steady state thermography measure the uninfluenced breast temperature, the dynamical one, first reduce the breast temperature with cooling in a desired time (usually between 2 and 6 minutes \cite{53_1}) on top of the breast and afterwards is measure the surface temperature. Nevertheless, parameters like cooling time, cooling temperature, general protocols and age, need further revision and validation. Besides, most of the studies remain in simulation phases \cite{12}. Kandlikar et al. \cite{12} review the main considerations regarding breast tumors simulation, like geometrical parameters, depth, size, and location of malignant or benign tumors. Lin et al. introduce a new methodology to simulate the early breast tumours using finite element thermal analysis; they considered parameters like temperature variance, breast contours, deepness, size and location of the tumour \cite{tem_new_6}.

\nopagebreak

%% file: Chapters/3_electrical_impedance.tex
% Third chapter of the Review

\section{Electrical Impedance Tomography}

Electrical Impedance Tomography (EIT) or Electrical Impedance Spectroscopy (EIS) is a technique used for evaluate the conductivity (also, permittivity) distribution inside an object (e.g. a breast) by measuring the voltages between electrodes located in a specific location. The procedure consists in applying a high frequency (between 50 and 100kHz) and low-current (less than 5mA) signal through electrodes in the skin. Identically, some electrodes record the voltage response in the skin, obtaining a "permittivity" factor. The electric conduction in a tissue can vary depending the type of tissue, the separation between electrodes, and significantly in the presence of cancerous tissue or tumor. As a matter of fact, Kubicek et al.  \cite{68} were one of the first researchers in use a four-band electrode (tetra-polar) configuration and EIT techniques to measure the cardiac output in 1970. The electrical impedance technique has been used for detecting several types of diseases such as cardiac arrest \cite{68_1}. Equally important, the features of EIT techniques play a key role in the breast cancer early-detection. The impedance of a living tissue is a complex number expressed by both, magnitude and phase, in fact, certain sub-features may come out after some pre-processing techniques, in order to reduce noise and make them convenient for MLT. Over the last decades, many research teams have disclosure some basic protocols for noise reduction and standardization purposes towards EIT such as frequency, max current, limiting circuit, room temperature, time of analysis, quantity of recorded signals (i.e., tetra-polar), impedance and input stray capacitance. Brown \cite{eit_new_1} provides a wider explanation on EIT for the health care area.

\subsection{Initial years of electrical impedance tomography}

The EIT systems for breast cancer diagnosis use digital tools to help physicians understand the electro-physical changes in the human body when a tumor or cancerous tissue exists. Kubicek et al. \cite{68} referenced the initial exploratory research of electrical impedance tomography. In the first place, Jossinet et al. \cite{22, 69} explains the main protocols for measure the body's electrical impedance; they use frequencies between 0.488kHz and 1MHz over twelve different points in many tissue samples. On the other hand, the features collected from each sample were (i) impeditivity at zero frequency (I0), (ii) phase angle at 500kHz, (iii) high-frequency slope of phase's angle, (iv) impedance distance between spectral ends (DA), (v) area under the spectrum, (vi) area normalized by DA, (vii) maximum value of the spectrum, (viii) distance between I0 and the real part of the maximum's frequency point and, (ix) the length of the spectral curve. Extra information regarding the technique, protocol, population and features are in \cite{22, 69}. Lastly, they used STATISTICA as analysis tool, which helped to create a set of rules based on features, obtaining an overall classification efficiency of 92\%. 

In 2003 Zou and Guo have reviewed some techniques regarding EIT for breast cancer detection, they suggest that is possible to determine whether is a malignant or benign tumor using the electrical breast properties. In particular, a breast with a malignant tumor has lower electrical impedance than the surrounding normal tissue \cite{70}. sMoreover, Zou cited a research article from 1926 \footnote{The journal of cancer research, AACR. Department of Biophysics, Cleveland Clinic Foundation, Ohio}, this study produced the first recorded ever of the electric capacity of a breast tumors (see \cite{71}). To summarize, the suspension of biological cells or a biological tissue when placed in a conductivity cell, behaves as though it were a pure resistance in parallel with a pure capacity. Certain types of malignant tumors have rather high capacity in comparison with benign tumors or with inactive tissues with similar characteristics, they concluded \cite{71}. Differently, Cheney et al. \cite{72} have proposed a Noser Algortihm \footnote{The inverse conductivity problem is the mathematical problem that must be solved in order for electrical impedance tomography systems to be able to make images \cite{72}} approach to solve the EIT reconstruction's problem. It provides a reconstruction with 496 degrees of freedom by images reconstructed from numerical and experimental data, including statistics from a human chest \cite{72}. In brief, the previously methodology helped other authors to develop better breast and tumor electric models. The next section describes the foremost technologies and MLT regarding EIT. 

\subsection{Computer aided techniques and electrical impedance tomography}

\begin{figure}[htb]
\centering
\includegraphics[width=5in]{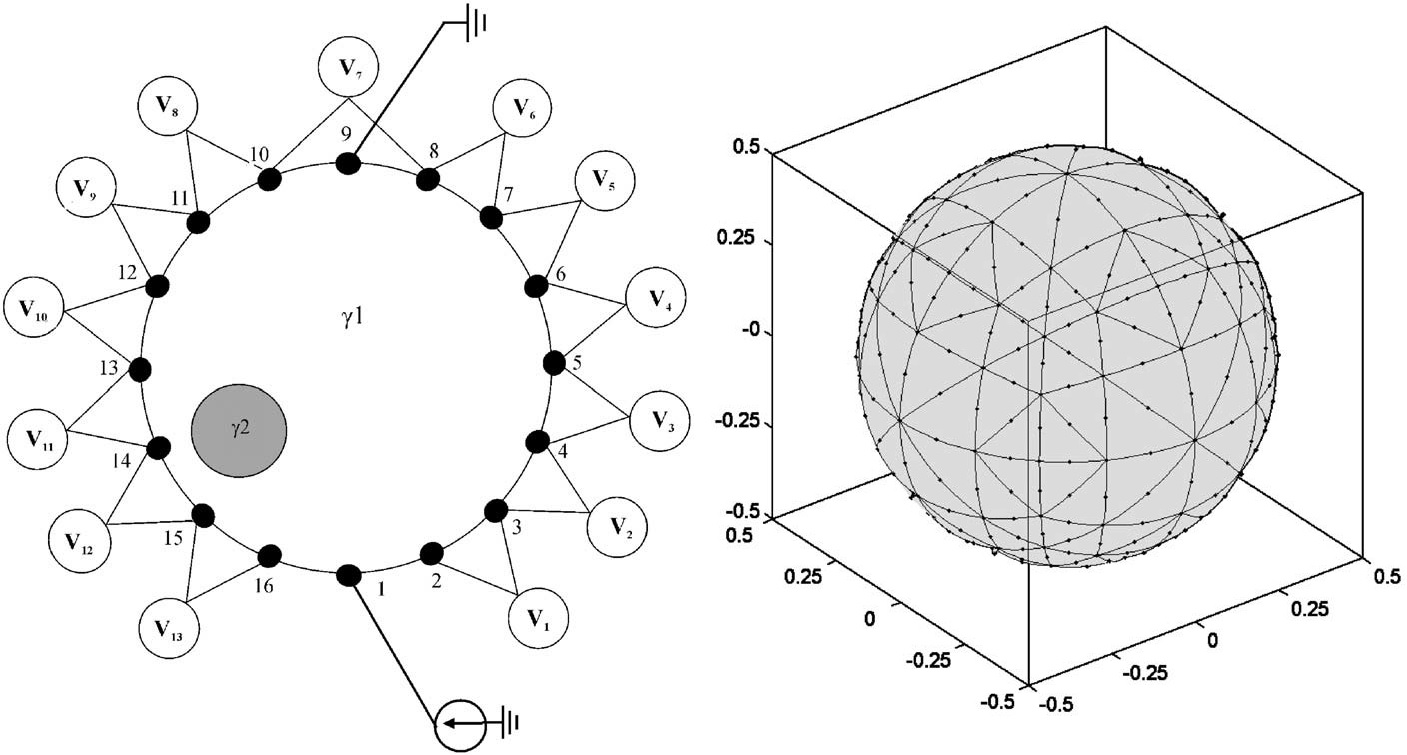}
\caption{(Left) electrode-to-electrode configuration; (right) discretization of the internal perturbation with isoparametric element in normalized dimensions from \cite{74}.}
\label{eit_3D_model}
\end{figure}

In 2007 Stasiak et al. \cite{74} presented a PCA analysis method together with neuronal networks for both localization and sizing of breast irregularities with EIT. It is important to recall that PCA is a statistical strategy for dimensionality reduction, used for transform a n-dimensional space into a smaller space, taking into consideration the possibly of correlation between variables or features. The PCA algorithm is normally applied in datasets before a MLT because, this algorithm reduce the quantity of features therefore, the overall computational cost (normally the accuracy decrease as well). As an illustration, The figure \ref{eit_3D_model} illustrate the sixteen-electrode arrangement on a typical EIT breast's test, where is applied a sinusoidal low current; after, it is measured the voltage differential between the electrodes. The Figure \ref{eit_3D_model} right’s side depict a tumor and the boundary interaction with the low current perturbations. Lastly, the simulated irregularity employs the boundary element method (BEM).

The ANN have been making a huge impact in pattern recognition on several biomedical signals in the last years, thus, Zheng et al. have made a study focused in resonance-frequency electrical impedance spectroscopy (REIS). With an initial set of 140 patients, including 56 who had biopsies, the performance of the overall system is evaluated with ANN and a case-based leave-one-out method \cite{75}. The team makes an equidistance arrangement of a seven electrode-probe on each patient, and then the data goes to a CAD system for interpretation purposes. In addition to ANN for EIT breast cancer prediction, Shetive, et al. \cite{76}  in 2012 create a multi-layer perceptron \footnote{A class of feedforward artificial neural network, at least is composed by 3 layers} (MLP) classifier model who achieved a 96\% accuracy on the Jossinet et al. database \cite{22, 69}. Similarly, logistic regression, KNN and Naive Bayesian networks were used by Calle-Alonso et al. \cite{77} to classify the EIT data set from \cite{22, 69}. Furthermore, the key point in obtaining a global accuracy of 97.5\% was transforming the possible six-classes breast tissue (or labels): (i) connective tissue, (ii) adipose tissue, (iii) glandular tissue, (iv) carcinoma, (v) fibroadenoma, and (vi) mastopathy, into two classes, (i) Carcinoma and (ii) Fib+Mas+Gla. In essence, the table 4: "Acc-1" refers to two-classes, "Acc-2" to three-classes and "Acc-3" to six-classes approach. 

\begin{table}[htb]
\tbl{Summarized Electrical Impedance Tomography. The main parameters of evaluation are: Accuracy (Acc), sensitivity (Sen), specificity (Sp), AUC (area under the curve), ROC (receiver operating characteristic curve), correlation coefficient (CC) and mean square error (MSE).}
{\small \begin{tabular}{p{7cm}p{3cm}p{2cm}c}
\toprule 
	Scope of the project &  Machine Learning \newline Technique (MLT) & Evaluation \newline Result &  Ref. \\
\midrule
	Impedance variability on six breast tissue, 9 different features, 1 target & No MLT & Mean ($\mu$) \newline Std. Dev. ($\sigma$)&\cite{22}\\ 

	Classification of breast sample tissues using STATISTICA\textregistered , 9 features, 1 target & Statistical analysis & Acc: 92\% &\cite{69}\\ 

	Modeling of EIT distribution in a breast with a tumor, using ANN, PCA amd BEM & ANN & Abs. E: 0.23 & \cite{74}\\ 

	Review of the main advances in EIT for breast cancer diagnosis & No MLT & -- &\cite{70}\\ 

	First test ever of EIT in breast cancerous tissue (1926) & No MLT & -- &\cite{71}\\
				 
	REIS system with 7 electrodes, 1 in the center, and 6 concentrically separated.REIS showed high false positive rate & ANN & Acc: 67\% \newline Sen: 54\% \newline Sp: 90\% & \cite{75}\\
		
	Multi-Layer perceptron algorithm for the EIT data set from \cite{22, 69} & ANN - MLP & Acc: 96\% \newline MSE: 0.1 \newline CC: 0.99 & \cite{76}\\
	
	CAD system for breast cancer classification in the EIT data set from \cite{22, 69} & LR + NBN & Acc-1: 97.5\% \newline  Acc-2: 89.7\% \newline  Acc-3: 77.35\% & \cite{77}\\
		
	EIT system for early detection of breast cancer in 1103 women & Multi LR & -- & \cite{78}\\

	10 women clinical study, using 2 different setups with a EIT-Probe & LDA, LSE & -- & \cite{79}\\		
		OCCII-GIC system for make a map of the breast, using 85 electrodes and frequencies from 10kHz to 3MHz & -- & -- & \cite{80}\\				
	\bottomrule
\end{tabular}}
\label{summarize_table_eit}
\end{table}

Given these points, advances in EIT have allowed the construction of different devices able to map and create an Electrical Impedance Map (EIM), in 2015 one team have use the \textit{T-Scan 2000ED}\footnote{T-Scan 2000ED, from Mirabel Medical Systems, Austin, TX} on a population of 1.103 women, identifying 29 cancers. Indeed, a multiple logistic regression analysis associate clinical variables and EIS results in \cite{78} study. Subsequently, Haeri et al.\footnote{Study from: Fraser Health Authority and Jim Pattison Outpatient Care and Surgery Centre (JPOCSC) with study number FHREB2014-065 and 2015s0156, respectively} \cite{79} divulge a clinical study using a two different EIT devices, the first setup, is composed by a Covidien electrodes, spectroscope HF2IS and trans-impedance amplifier HF2TA from Zurich Instruments. The second setup, is EIS-Probe similar to the first one, but its electrodes and their location of installation are different, using least absolute deviation (LAD) and least square method (LSM) are implement for data's analysis. Equally important, Zarafshani et al. \cite{80} propose a 85 electrodes board to create an Electrical Impedance Mammogram similar to an EIM, likewise, the main device is describe as a wide bandwidth EIM system using novel second generation current conveyor operational amplifiers based on a gyrator (OCCII-GIC) and the input current range from 10kHz to 3MHz. Table \ref{summarize_table_eit} describes references regarding electrical impedance technologies. The next section will present the main EIT devices, nevertheless, mostly of them are not available o the market. 

\subsection{Main electrical impedance tomography devices}

During the last years, many researchers have been struggling to standardized the EIT test for breast cancer. This review will be focused in five devices, where the main differences are: (i) physical such as the number of electrodes, where range between 64 and 256, the method of measurement range from but no limited to the one where the patient is lying on the bed, a probe that a human-expert managed or a wearable bra. (ii) The electrical part includes the frequency and magnitude of the low-current signal, the electronic components, and the minimum detectable tumor’s size. In general, these devices made part of a CAD system compose by an expert and algorithm that give insights of the probability of developing cancer. Likewise, each year more authors explain the advantages of combining CAD systems with human experts, changing the one-step into two-steps diagnosis systems. The EIT devices available on the market and research area are in Table \ref{table_eit_devices}. A recent study use the bioimpedance analyzer MScan1.0B in 489 patients obtaining parameters as a function of permittivity and conductivity behavior on the breast, the Sn (92.4\%), Sp (96.0\%) results demonstrated the EIT feasibility and low-cost \cite{81}. Given these points, Feza et al. \cite{86} and Lima et al. \cite{87} present two models of electro-thermal system for medical diagnosis. They have concluded that the improvement on the accuracy is greatly augmented when the techniques are employed the same time, rather than performing the diagnostic separately. Likewise, Singh et al. \cite{eit_new_2} build a general purpose EIT device that uses multifrequency electrical impedance in medical imaging; they explained the main features are resolution, current amplitude, number of electrodes, size and range of frequency.\\

"Insert table \ref{table_eit_devices} here (the table must be in landscape mode, therefore, it should be let on appendices)" \\

The last advances in MLT and miniaturization have helped to create more accurate and robust system. The Table \ref{table_eit_devices} shows fice devices for EIT analysis. Those devices may vary depending on the size; nevertheless, the EIT-BRA proposed by \cite{19} is a huge advance in wearable devices for detect breast cancer. The next chapter reviews the most up to date system mixing both techniques, thermography and EIT.

%% file: Chapters/4_electro_thermal.tex
\section{Electrical impedance tomography and thermography combined systems}

\begin{figure}[htb]
\centering
\includegraphics[width=4in]{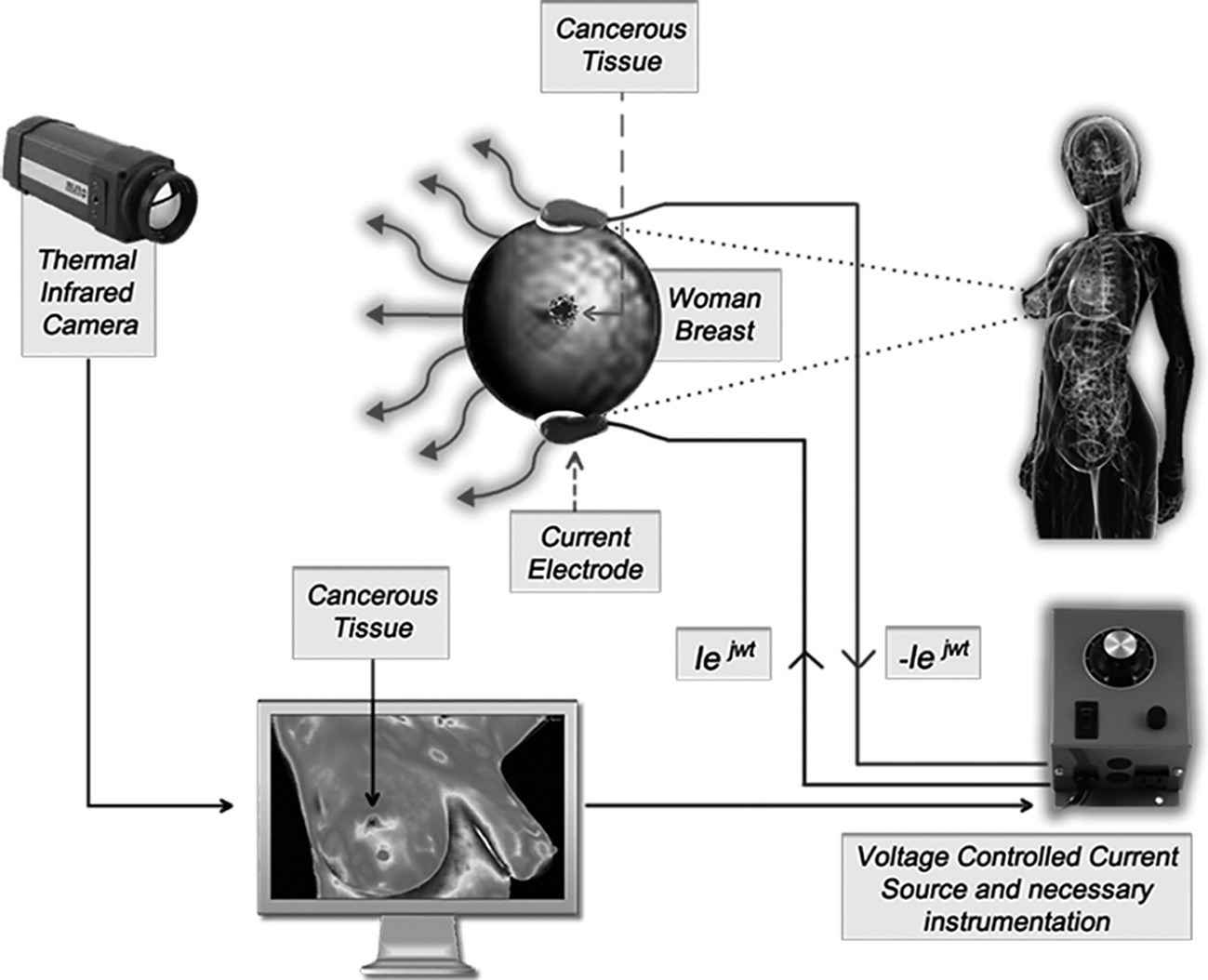}
\caption{Block diagram of Feza et al. Electro-Thermal Imaging System, CAD and EIT system with IR camera \cite{86}.}
\label{eit_thermal_1}
\end{figure}

During the last decade, several authors present different works on electro-thermal architectures for breast cancer diagnosis. The electro-thermal word refers to systems made of both techniques EIT and thermography. Feza et al. suggest that a hybrid system is needed in order to improve the performance of breast carcinoma diagnosis; in fact, each technique has weaknesses that are highly reduce in electro-thermal systems. The proposed method provides a better contrast resolution, in other words, tumors between 3mm and 9mm were saw with this CAD technique. In general, most of the studies just target the methodology, then, it is needed practical implementation. An electro-thermal system works as follows, first a low current is injected on the breast, then, an IR camera takes a snapshot of the breast. Under those circumstances, what will be the difference? In detail, the cancerous tissue has almost five to ten-times larger electrical conductivity factor than normal tissue, for that reason, the breast heat map will change and show other insights on the final image. The current’s frequency may change the results, for that reason in \cite{86}, different parameters are tested. The Feza et al. system works as presented in Figure \ref{eit_thermal_1}, firstly, an electrical current goes through surface of the breast, controlling both, voltage and current. Secondly, an IR camera capture the breast surface temperature. Finally, a database is then created that will fed a CAD system coupled with one or several MLTs.

A recent electro-thermal study from Menegaz and Guimaraes explains that some natural or unnatural (others diseases) processes in the human body, can lead to notably temperature gradients, as a result the thermography evaluation could give erroneous outcomes. They explain the impedance as the ratio between the variations of the structure’s surface temperature and the external modulated heat flux. On the other hand, they validate the method with silicone phantom samples using hyperplastic materials with simple geometry. A damage metrics or “cancerous tissues” with different thickness measure the global performance in \cite{87}. Nonetheless, is known that phantom tissue is commonly used for test new bio-electrical-based devices, but surely, real patients will provide better insights.

%% file: Chapters/5_discussion_conclussion.tex
\section{Discussion}

The early-detection of BC plays a key role in reducing the mortality rate; nevertheless, many authors explain the limitations when only humans take part in the breast cancer judgement, occasionally with non-permissible false positive and false negative cases. The integration of CAD systems into new BC screening methods like EIT and thermography, surely will boost the global performance. In addition, higher the confident rate higher the technique cost (as showed Table 1), therefore, many researchers have been working in develop new easy-to-use, intelligent and inexpensive systems for detecting BC. Indeed, these objectives allowed the inclusion of MLT in the pipeline for BC detection; Yassin et al. \cite{56} review the MLT in medical imaging for BC, explaining that the biomedical information like image, signals or stacked data, are a complex set of information that describe a person health. To emphasize, in many cases for a physician is hard to be completely sure in the diagnostic, likewise, many features are hidden to “human eyes” even in experts. Under those circumstances, intelligent systems are capable of “see” these out of sight features that may perhaps increase the performance. Normally, a diagnosis system is made of (i) a measurement phase (e.g. thermography, EIT, mammography, MRI), (ii) a CAD system will pre-process, find hidden patterns, post-process and diagnose (the process may change depending on the measurement technique)a (iii) finally, a patient will get a result based on both, physician and the CAD system.

On the other hand, the data itself lies as main drawback in many ML pipelines, for example, Sathish et al. \cite{59} have a 20\% accuracy increase in BC thermography diagnosis when feature engineering is applied (min-max normalization of thermal images). The feature engineering tries to pre-process the input data in a way that the computer can distinguish easier the target disease, for example, normalization, ROI segmentation, filters, polynomial features and dataset expansion, data augmentation, and so forth. In contrast, the cost, accuracy and wearability of the system are crucial factors to address. Under those circumstances two questions arise; firstly, is it the system suitable for high, middle or low-income communities? Secondly, it is sufficiently autonomous and robust the CAD system to accurately detect BC? Consequently, the Table 1 gives a starting point to assess those questions. This section deals with two significant points.

Firstly, the thermography has grown the last years as an important technique for BC diagnosis; in particular, a greater number of studies have been targeting the only freely available database on the web \cite{51, 52} for thermal images. Unfortunately, opposed to other types of bio-medical databases (e.g., mammography, cancer cell images, ultrasound, MRI) there is not many databases for thermography, aside the one above-mentioned. The best results across the reviewed MLT are ANN, DNN \cite{2, 59, 65} up to 98\%, SVM \cite{2, 50, 62, 65, 66} up to 91\% and RF \cite{59} equal to 95\% in accuracy, (see Table 2). It is important to realize that ANN models like CNN or DNN are more robust than a typical LR, SVM or RF. Artificial neural networks have the ability to identify hidden patterns that others models cannot. To point out, a CNN is made of units called “neurons”, layers and connection between layers; thence, higher the quantity of layer, higher the model’s deepness. Nonetheless, all the MLT have to deal with problems such as under-fitting, over-fitting, hyper-parameters selection, architecture, banishing and exploding gradients \cite{56, 57}. Borchartt et al. \cite{57} argue that the importance of feature extraction and interpretation are vital steps to achieve the desired performance, also they characterize the main public and private database and exhort the community to create additional ones, in order to support research teams. 

Secondly, EIT is also a technique capable of detect BC however, fewer studies exist due the lack of public databases \cite{22, 69}. A part of the researchers are focusing in developing (i) 3D models with diverse electrical tissue properties \cite{74}, another part in (ii) portable and non-portable devices \cite{19, 81, 84, 85, 86, 87} as shown in Table 6. Furthermore, the phenomenon and performance with CAD systems are reviewed \cite{68, 69, 70, 71, 72, 73, 75, 76, 77, 78, 79, 80}. Then, two teams have demonstrated a considerable reduction in FP and FN rates in electro-thermal CAD systems \cite{23, 83}. To put it differently, the miniaturization of electronic devices and the hyper-connection of a globalized world suggest that in the short-term future, many new wearable and portable devices will track our health day-to-day. Consequently, wireless and wearable devices will track breast health constantly, reducing the BC mortality rates. Nevertheless, it stills a debate about whether the data structure could influence the performance, in detail, some EIT + CAD systems map the breast's bio-impedance creating an “image” \cite{19, 84, 87}, on the other hand, several systems prefer to obtain frequency-based parameters \cite{22, 23, 69, 81}, like resistivity, phase angle, and so forth (see section 3). Finally, some questions still unanswered, which EIT approach is better, image-centered or parameter-centered? Will be in the near future a standard protocol for EIT BC diagnosis? 

Thirdly, the BC might affect whomever woman, but specific considerations, like gene expression, nutrition habits, hereditary issues and dense breast may increase the chances of BC develop. Indeed, new techniques like DNA (deoxyribonucleic acid) gene expression are tested for cancer diagnosis like breast, bladder \cite{88}, leukemia \cite{89}, human colorectal carcinoma \cite{90}, prostate \cite{91}, breast \cite{92}, and so forth. Even though, these techniques remain highly expensive for the middle and low-income patients. 

Although the above paragraphs express the author’s opinion, many questions still pending, e.g. is it abundantly and well balance the available databases? Is it necessary to follow a protocol prior to an EIT or thermal test? Is it enough the FP and FN rate or the current models still lack performance? Those questions may bring new insights in the future work that researchers would take. It is important to recall that we  presented the importance of machine learning techniques in the CAD systems. On the contrary, the lack of public databases is a huge problem that has been limiting the research outcomes in the two-discussed techniques for BC detection. Lastly, the main purpose of this review is (i) to give insights to the readers about the current advances in thermography and EIT, for early-BC detection; but also, (ii) to point out that thermography and EIT do not pretend to replace mammography, but rather become in a prior techniques that being cheap and easy-to-implement will be accessible for underdevelopment countries. 

\section{Conclusion}

\textbf{Thermography and EIT are not contemporary techniques for breast cancer screening, but they were not considered as parallel techniques before the last decade, when the increase in thermal cameras performance and the advances in computing brought to light machine learning algorithms and CAD systems capable of aid physicians to interpret in a better way bio-medical databases, consequently reducing the FP and FN rates. We would like to emphasize that these techniques target low-cost, affordable and radiation-free CAD systems for diagnose breast cancer, rather than the common standard, mammography. The state-of-the-art involves the background, technical details, application in CAD + MLT systems and last advances in thermography and EIT. The review explains the most popular MLTs, nevertheless, each year new methods, techniques and architectures become known. Consequently, it is difficult to compare the performance between several studies in a different time-frame. Despite the advantages of CAD systems, clinically, some countries present problem with the high rate of FP and FN, even when MLT are applied, thus, many systems need revision and rearrangement in how they are implemented. 
}

As mentioned through this review, there is global scarcity in public clinical databases. Furthermore, it is important to mention the benefits of publicity and standardization regarding clinical databases, where those could help research teams in finding resources, create reliable links and develop new and robust models. We advise that future systems should by made up of two or more different databases, for example electro-thermal, electro-thermal-mammography systems, where a patient’s information “not directly related” with breast cancer could support the global system. New ANN techniques like RNN and self-normalizing networks, auto-encoders, gradient boosting machines, or optimization methods like particle swarm, parzen tree, evolutionary algorithms, artificial bee colony and so forth, surely will increase the model performance giving the CAD systems the needed reliability for a wide implementation either, in hospital or wearable devices. Therefore, the portable and wearable devices are promising trends that appeared in the last years, thus, developing reliable devices and machine learning libraries capable of measure the chance of having cancer, without the need of qualified personal and low-cost, will influence widely the research community and the global population. At the same time, is important to have in mind that the key goal is not to remove the physicians from the diagnosis phase, but strengthen the current methodologies. Finally, new programming libraries for machine learning developing, like Scikit-Learn \cite{93} and TensorFlow/Keras \cite{94} have been growing not just in quantity of MLT but also in optimization and robustness making them feasible as CAD systems.

%% file: Chapters/6_Appendices.tex
% setting the counter for the new tables 
\setcounter{table}{0}
\begin{sidewaystable}
\tbl{\large Comparison of breast cancer screening and diagnosis techniques, structured from \cite{12}.}
{\small \begin{tabular}{p{3cm}p{2.5cm}p{1.4cm}p{1.4cm}p{1.4cm}p{3.5cm}p{1.4cm}p{3.5cm}p{3.5cm}} \toprule
 & \multicolumn{2}{l}{Characteristics} \\ \cmidrule{2-9}
Technique & Mechanism of operation & Sensitivity & Specificity & Cost & Method & Wearable & Cause of discomfort & Recommend for \\ 
\midrule
Mammography & Low energy X-rays & 90\% & \textgreater{}94\% & Moderate & Compressed the breast & No & Pain in the breast & Screening and diagnostic \\

Magnetic Resonance Imaging (MRI) & Magnetic field and pulsating radio waves & 90\% & 50\% & High & Contrast substance injected and dynamic images obtained & No & Claustrophobia, reaction to contrast agent, renal insufficiency patients & Screening in women at high risk for breast caner \\

Positron Emission Tomography (PET) & Gamma rays emitted by tracer substance & 90\% & 86\% & High & Small amount of radioactive tracer injected in the body & No & No & Determine if cancer has spread to other part of the body \\

Ultrasound & High frequency sound waves & 82\% & 84\% & Low & Hand-held or automated ultrasound device & No & No & Screening in dense breast \\

Tomosynthesis (3D Mammography) & Low energy X-rays & 84\% & 92\% & Low & Compressed the breast & No & Pain in the breast & Screening and diagnostic \\

Electronic Palpation Imaging (EPI) & Pressure changes & 84\% & 82\% & Low & Hand-held electronic, tactile sensor & Possible & Pain in the breast & Follow-up after abnormal findings \\

Thermography & Surface Temperature measurement & \textgreater{}90\% & \textgreater{}90\% & Low & Temperature sensors attached to the skin’s surface & Yes & No & Screening \\

Electrical Impedance Tomography (EIT) & Electrical Impedance in the tissue & 87\% & 82\% & Low & Electrodes attached to the skin’s surface & Yes & Tickling for current variation & Screening \\

Biomarkers from Blood Sample Test & Blood samples biomarker & 82\% - 88\% & 85\% - 90\% & Low & Blood results and interpretation & No, test in situ & No & Screening \\ 

\bottomrule
\end{tabular}}
\label{x_comparison_between_technique}
\end{sidewaystable}

\setcounter{table}{4}

\begin{sidewaystable}
\tbl{\large Electrical Impedance Tomography devices and properties.}
{\large
\begin{tabular}{p{3cm}  p{3.5cm} p{3.5cm}  p{3.5cm}  p{3.5cm}  p{3.5cm}}
\toprule
	   Reference & USA \cite{23} & Russia \cite{83} & Germany \cite{84} & Korea \cite{19} & USA \cite{85} \\
	   \midrule
	   
	   Device &
	   \includegraphics[width=1.45in]{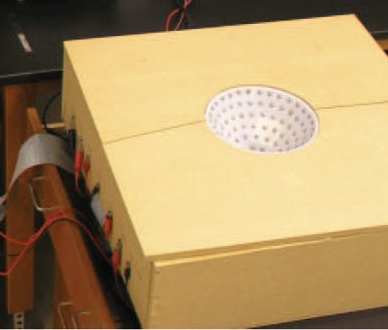} &
	   \includegraphics[width=1.45in]{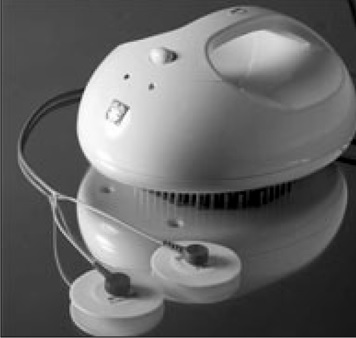} &
	   \includegraphics[width=1.45in]{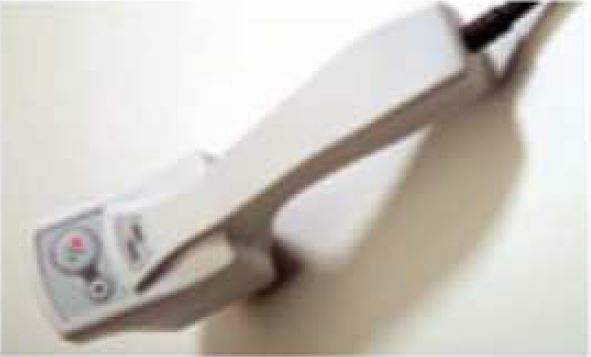} &
	   \includegraphics[width=1.45in]{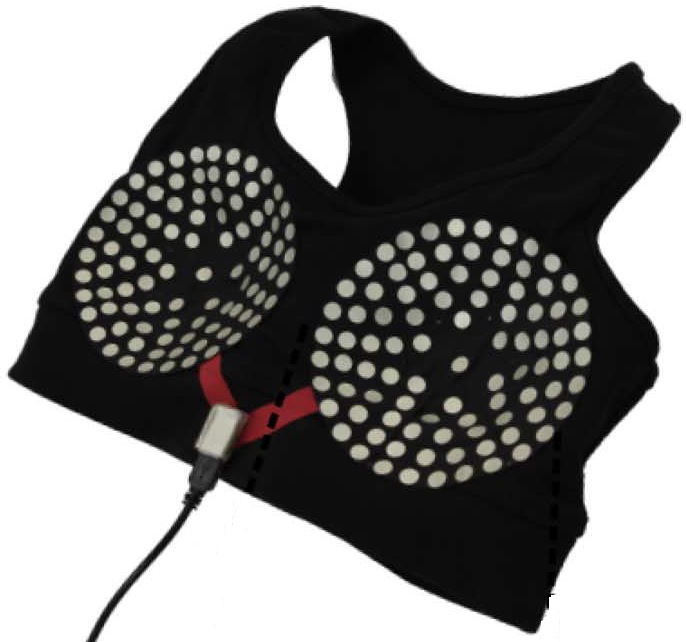} &
	   \includegraphics[width=1.45in]{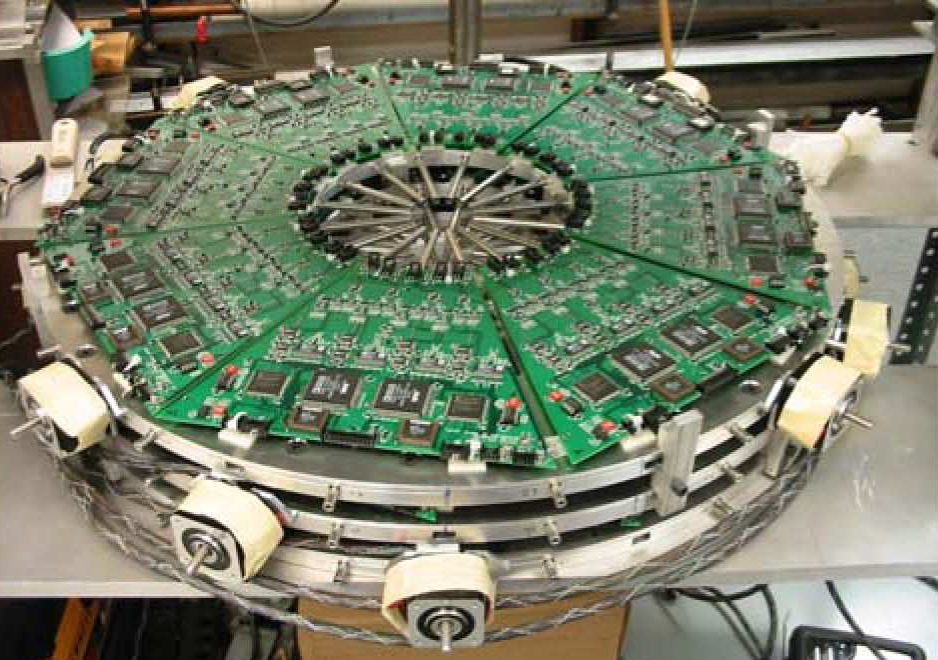} \\
	   
	   Year & 2008 & 2012 & 2001 & 2014 & 2007 \\
	   Location & Bed & Hand-held device \newline +ref. electrode & Probe \newline + ref. Probe & BRA (wearable) & Bed \\
	   Dimension & 11.7 cm height \newline 19.1 cm diameter & 16x18x10 cm & 7.2x7.2 cm & 30x25x5 cm & 60 cm diameter \\
	   Weight & N/A & 2 Kg & N/A & 72g & N/A \\
	   Img. device & Computer & Computer & Computer & Mobile device & Computer \\
	   Dimension & 2D slices & 2D slices & 2D  & 3D & 3D \\
	   \# electrodes & 128 (7 layers) & 256 (planar) & 256 (planar)  &  92 (flexible) & 64 (4 layers) \\
	   Frequency & 10kHz & 10 - 50 kHz & 58 Hz - 5kHz  &  100 Hz - 100kHz & 10kHz - 10MHz \\
	   Amplitude & 1mA & 0.5mA & 1V - 2.5V & 10$\mu$A - 400$\mu$A &  N/A \\
	   SNR (dB) & 77dB & N/A & N/A  &  90dB & 94dB \\
	   Minimum \newline detectable size  & 12mm & N/A & N/A  &  5mm & N/A \\
	   \bottomrule	   	   	   
	\end{tabular}}
	\label{table_eit_devices}
\end{sidewaystable}